\documentclass[letterpaper,10pt]{article}

\usepackage{amsmath}
\usepackage{amssymb}
\usepackage{psfrag}
\usepackage[dvips]{graphicx}
\usepackage{pdfpages}

\newcommand{\dirac}{{\slash \negthinspace \negthinspace \negthinspace \nabla}}

\newcommand{\chypergf}{{}_{1}F_{1}}

\author{A.\ L\'opez-Ortega\thanks{alopezo@ipn.mx}, M.\ Y.\ Rojas-Luis \\ 
Departamento de F\'{\i}sica. Escuela Superior de F\'{\i}sica y Matem\'aticas. \\
Instituto Polit\'ecnico Nacional. \\
Unidad Profesional Adolfo L\'opez Mateos. Edificio 9. \\
M\'exico, D.\ F., M\'exico. \\
C.\ P.\ 07738 }

\title{Late time behavior of the massive Dirac field in $D$-dimensional Minkowski spacetime}

\begin{document}

\maketitle

\begin{abstract}

To extend previous results on the late time behavior of massive fields, for the Dirac field propagating in the $D$-dimensional Minkowski spacetime we calculate analytically its asymptotic tails. We find that the massive Dirac field has an oscillatory inverse power law tail. The frequency of the oscillations depends on the mass of the field and the power law decay rate depends on the dimension of the spacetime and the mode number of the angular eigenvalues. We also compare with previous results in curved spacetimes. 

KEYWORDS: Massive Dirac field; $D$-dimensional Minkowski spacetime; Late time behavior

PACS: 04.30.Nk; 04.20.Ex; 04.50.-h
 
\end{abstract}

\section{Introduction}
\label{s: Introduction}

At present time, it is well established that the evolution of a test field around a black hole has three stages \cite{Leaver:1986gd}, \cite{Berti:2009kk}, \cite{Konoplya:2011qq}: a) Prompt response, b) Quasinormal ringing, c) Asymptotic tail decay. Furthermore, it is well known that each stage is associated to a different part of the contour in the inversion formula for the Green function. Thus we know that the quasinormal ringing comes from the poles, the asymptotic tail decay is associated with the branch cut, and the prompt response comes from the semicircle with large radius \cite{Leaver:1986gd}, \cite{Berti:2009kk}, \cite{Konoplya:2011qq}.

Beginning with the work by Price \cite{Price:1971fb}, \cite{Price:1972pw} has been interest in determining the asymptotic tails in several spacetimes  \cite{Leaver:1986gd}, \cite{Konoplya:2011qq}, \cite{Ching:1994bd}, \cite{Ching:1995tj}. The previous studies are focused on massless bosonic fields, as the gravitational, Klein-Gordon, and  electromagnetic perturbations \cite{Konoplya:2011qq}--\cite{Ching:1995tj}. It is well known that the late time behavior of the massive fields is different from that of the massless fields and for example the massive Klein Gordon field has tails in Minkowski spacetime \cite{Morse-Feshbach} since the different frequencies forming a massive wave packet propagate  with different phase velocities. For massive fields propagating in curved spacetimes we find that the computations of the late time tails are focused on the Klein-Gordon field (see \cite{Hod:1998ra}--\cite{Koyama:2001qw}), although the asymptotic tails of the massive Dirac field have been calculated, mainly in four-dimensional spacetimes \cite{Jing:2004xv}--\cite{Moderski:2008nq} (see \cite{Hod:2013dka} for the massive spin 2 field and \cite{Konoplya:2006gq} for the massive vector field). Furthermore we know calculations of the late time behavior for massless fields \cite{Cardoso:2003jf} and for the massive Klein-Gordon field \cite{Moderski:2005hf} in higher dimensional spacetimes.

Also the different aspects of fermion's propagation  in spacetimes have not been extensively explored, as for the  bosonic fields. Thus it is convenient to study in detail the  propagation of fermionic fields in different spacetimes. Therefore, in this work we analytically calculate the asymptotic tails of the massive Dirac field propagating in the $D$-dimensional Minkowski spacetime ($D \geq 4$), in order to extend the results that appear in Refs.\ \cite{Jing:2004xv}--\cite{Moderski:2008nq} and in the previous studies on the propagation of the Dirac field in higher dimensional spacetimes \cite{LopezOrtega:2009zx}--\cite{Chakrabarti:2008xz}.

Owing to the problem of analyzing the dynamics of massive fields is more challenging than the corresponding problem for massless fields we calculate the late time tails of the massive Dirac field in $D$-dimensional Minkowski spacetime since we like to know how the initial configuration of this field evolves in the time and in the previous results we find that as $t \to \infty$ a massive field may behave qualitatively different from a massless field. Also we analyze this physical situation since calculating the asymptotic time tails of the massive Dirac field in $D$-dimensional Minkowski spacetime allows us to separate the effects of field's mass and the curvature of spacetime. Thus we can study only the effect of the mass on the asymptotic time behavior, that is, this setting allows us to clarify what effects are caused by the mass of the field in the time evolution when the spacetime curvature is absent.

For massless fields we know that the power law fall-off is sensitive to the fact that the spacetime dimension is odd or even \cite{Cardoso:2003jf}, \cite{Soodak-Tiersten}. We think that it is relevant to determine whether the mathematical form of the asymptotic tails of the massive Dirac field depend on whether $D$ is odd or even. Furthermore our computation allows us to analyze in a simple setting the influence of the spacetime dimension in the late time behavior of this field. Therefore, the computation of these characteristics about the late time tails of the massive Dirac field extend our knowledge on its behavior and can be useful when more complex higher dimensional spacetimes will be studied, since we expect a richer phenomenology in $D > 4$ and our analysis can help us to understand what aspects of the relaxation phenomenon depend on the existence of a body that produces curvature and what properties depend on the spacetime dimension and the mass of the field.

We organize the rest of this paper as follows. In Sect.\ \ref{s: Tails Dirac} we calculate the late time tails of the massive Dirac field propagating in the $D$-dimensional Minkowski spacetime. First we determine the late time tails for $D$ even in Subsect.\ \ref{ss: D even} and then for $D$ odd in Subsect.\ \ref{ss: D odd}. Finally in Sect.\ \ref{s: Discussion} we discuss our main results.

\section{Tails of the Dirac field}
\label{s: Tails Dirac}

In the coordinates ($t$,$r$,$\varphi_i$), with $i=1,2,\dots,D-2$,  a $D$-dimensional maximally symmetric spacetime has a line element 
\begin{equation}\label{e: MetricaDiracES}
ds^{ 2 } = E(r)^{ 2 }dt^{ 2 }- Q(r)^{ 2 } dr^{ 2 }-H(r)^{ 2 }d{ \Sigma}_{ D-2 }^{ 2 },
\end{equation} 
where $E$, $Q$, $H$ are functions of the coordinate $r$ and $d{ \Sigma   }_{ D-2 }^{ 2 }$ is the line element of the $(D-2)$-dimensional invariant base manifold with coordinates $\varphi_i$. To study the behavior of the spin $1/2$ field in spacetimes of the form (\ref{e: MetricaDiracES}) we use that the Dirac equation $i \dirac \psi = m \psi $ simplifies to the coupled system of two partial differential equations \cite{Gibbons:1993hg}--\cite{Cotaescu:1998ay} (see also Ref.\ \cite{LopezOrtega:2009qc})
\begin{align} \label{e: EcuacionDiracSimplificado}
 \partial_{ t }\psi _{ 1 }+\frac{ E }{ Q}  \partial_{ r }\psi _{ 1 }  =-\left(i\kappa \frac{ E}{ H } +i m E\right)\psi_{ 2 }, \\
 \partial_{ t }\psi_{ 2 }-\frac{ E }{ Q }  \partial_{ r }\psi _{ 2 }  =\left(i\kappa \frac{ E }{ H } -i m E\right)\psi_{ 1 },\nonumber
\end{align}
where $\psi_1$, $\psi_2$ are the components of a two spinor and $\kappa$ are the eigenvalues of the Dirac operator on the base manifold with line element $d{ \Sigma}_{ D-2 }^{ 2 }$.

For the $D$-dimensional Minkowski spacetime, whose metric can be written in the form (\ref{e: MetricaDiracES}) with $E=Q=1$, $H=r$, and $d{ \Sigma   }_{ D-2 }^{ 2 }$ being the line element of the $(D-2)$-dimensional sphere, we obtain that the system of equations (\ref{e: EcuacionDiracSimplificado}) transforms into
\begin{align} \label{e: ParcialDiracMinkoswki} 
 \partial_{ t }\psi _{ 2 }-\partial_{ r }\psi_{ 2 }  =\left(\frac{ i\kappa  }{ r } -im\right)\psi_{ 1 }, \qquad
  \partial_{ t }\psi_{ 1 }+ \partial_{ r } \psi _{ 1 }  =-\left(\frac{ i\kappa  }{ r } +im \right)\psi _{ 2 },  
\end{align}  
with $\kappa =  \pm i ( n + (D-2)/2 ) $ representing the eigenvalues of the Dirac operator on the $(D-2)$-dimensional sphere where the parameter $n$ takes the values $n=0,1,2,\dots$, \cite{Camporesi:1995fb}. In what follows we choose $\kappa =  i (n + (D-2)/2)$. We think that similar conclusions are obtained for the other eigenvalues.

Defining $F_1$ and $F_2$ by
\begin{align} \label{e: FuncionF1F2}
 F_{ 1 }=\psi_{ 1 }+ \psi_{ 2 },  \qquad \qquad
 F_{ 2 }=\psi_{ 1 }- \psi_{ 2 },
\end{align}
from Eqs.\ (\ref{e: ParcialDiracMinkoswki}) we find that these functions satisfy 
\begin{align}\label{e: sumrest}
\left(\partial_r-\frac{ i\kappa  }{ r } \right) F_{ 2 }=-(im+\partial_t )F_{ 1 }, \qquad
\left(\partial_r+\frac{ i\kappa  }{ r } \right)  F_{ 1 }=(im-\partial_t )F_{ 2 }.
\end{align}
The differential operators that appear in the previous equations satisfy
\begin{equation}
 \left( \partial_r \pm \frac{i \kappa}{r} \right) \left( \mp im - \partial_t \right) \Phi = \left( \mp im - \partial_t \right) \left( \partial_r \pm \frac{i \kappa}{r} \right) \Phi .
\end{equation} 
Taking into account this fact, from Eqs.\ (\ref{e: sumrest}) we can obtain decoupled equations for the functions $F_1$ and $F_2$. The equations take the common form\footnote{In what follows $F$ represents to $F_1$ or $F_2$.}
\begin{equation} \label{e: common form F}
 { \partial  }_{ r }^{ 2 } F -{ \partial  }_{ t }^{ 2 }F - \left(\frac{L(L+1) }{r^2} + { m }^{ 2 } \right) F = 0,
\end{equation} 
with 
\begin{equation} \label{e: L for F1 F2}
 L=n + \frac{D}{2}-2 , \qquad \quad \left( L=n + \frac{D}{2}-1 \right),
\end{equation} 
for the function $F_1$ ($F_2$). Furthermore we point out that in both cases the quantity $L$ is an integer for $D$ even and it is a half-integer for $D$ odd.

To solve Eq.\ (\ref{e: common form F}) we use the Green function $G(r,r^\prime;t)$ that satisfies \cite{Ching:1994bd}, \cite{Ching:1995tj}
\begin{equation}\label{e: GreenRetardada}
 \partial_r^{ 2 }  G   - \partial_t^{ 2 } G    - \left(\frac{L(L+1) }{r^2} + { m }^{ 2 } \right)  G =\delta (t-t^\prime)\delta (r-r^\prime) .
\end{equation}
The causality condition imposes that for $t \leq 0$, $ G(r,r^\prime;t) = 0$. To find Green's function we follow the usual path and use its Fourier transform  \cite{Ching:1994bd}, \cite{Ching:1995tj}
\begin{equation}\label{e: FourierGreen}
\tilde{ G } (r,r^\prime;\omega) =\int _{ 0 }^{ \infty  }{ G (r,r^\prime;t) \textrm{e}^{ i\omega t }dt } ,
\end{equation} 
that is a solution of the differential equation 
\begin{equation} \label{e: Green tilde equation}
 \frac{ d ^{ 2 } \tilde{G}}{ dr^{ 2 } }   + \omega^{ 2 } \tilde{ G }  -  \left(\frac{L(L+1) }{r^2} + m^{ 2 } \right) \tilde{ G } =\delta (r-r^\prime) .
\end{equation} 
As is well known, $\tilde { G } (r,r^\prime;\omega)$ is given by  \cite{Ching:1994bd}, \cite{Ching:1995tj}
\begin{equation} \label{e: Fourier transform}
 \tilde{G} (r,r^\prime;\omega) = \frac{f_a(r^\prime,\bar{\omega}) g_b (r,\bar{\omega}) }{ W(\bar{\omega}) } ,
\end{equation} 
where $f_a(r^\prime,\bar{\omega})$ ($g_b (r,\bar{\omega})$) is the solution to the homogeneous form of Eq.\ (\ref{e: Green tilde equation}) that satisfies the left (right) boundary condition and $W(\bar{\omega})$  is the Wronskian of these solutions. To determine  $G(r,r^\prime;t)$ from $ \tilde{G} (r,r^\prime;\omega)$ we use the inversion formula  \cite{Ching:1994bd}, \cite{Ching:1995tj}
\begin{align}\label{e: Green}
 G(r,r^\prime;t)=\frac{ 1 }{ 2\pi  } \int _{ -\infty +i\delta }^{ +\infty +i \delta }{ \tilde{ G } (r,r^\prime;\omega) \textrm{e}^{ -i\omega t }d\omega } ,
\end{align}
where $\delta$ is a small positive constant. 

To solve the homogeneous form of Eq.\ (\ref{e: Green tilde equation}) we propose that its solutions take the form \cite{Hod:1998ra}, \cite{Hod:1997fy}
 \begin{align}\label{eq: Propuestaf}
 F  = r^{ L+1 } \textrm{e}^{ -\bar{\omega} r}   \phi (r)  ,
\end{align} 
with $\bar{\omega} = \sqrt{m^2 - \omega^2 }$, to find that the function $\phi$ must be a solution of the differential equation 
\begin{align} \label{e: radial phi}
z \frac{ d ^{ 2 }\phi  }{ dz^{ 2 } } +\left[ 2(L+1)- z \right] \frac{ d\phi  }{ dz } -(L+1)\phi =0,
\end{align}
where we use the variable $z = 2 \bar{\omega} r$. The previous equation is of confluent hypergeometric type \cite{Abramowitz-book}, \cite{NIST-book}
\begin{align}\label{e: EcuacionHipergeometricaConfluente}
z \frac{d^2 h}{d z^2} +[\gamma -z] \frac{d h}{d z} -\alpha  h=0,
\end{align}
with parameters $\alpha=L+1$, and $\gamma =2(L+1) = 2 \alpha$.

In $D$-dimensional Minkowski spacetime to calculate the tails of the massive Dirac field we impose the boundary conditions
\begin{description}
 \item[i)] The field goes to zero at $r=0$
 \item[ii)] The field is purely outgoing (decaying) for $\omega > m$ ($m > \omega$)
\end{description}
Considering the behavior in different regions of the solutions to the confluent hypergeometric differential equation  we find that the function satisfying i) is
\begin{equation} \label{e: left solution}
 f_a(r,\bar{\omega})  = A r^{ L+1 }\textrm{e}^{ - \bar{ \omega}  r } \chypergf (\alpha ,\gamma; 2 \bar{\omega} r) ,
\end{equation} 
where $\chypergf (\alpha ,\gamma;z) $ denotes the confluent hypergeometric function \cite{Abramowitz-book}, \cite{NIST-book}, and the function that satisfies ii) is equal to 
\begin{equation}  \label{e: right solution}
 g_b (r,\bar{\omega}) = B r^{ L+1 }\textrm{e}^{ - \bar{\omega}  r } U(\alpha ,\gamma; 2 \bar{\omega} r) ,
\end{equation} 
where $U(\alpha ,\gamma;z)$ is the Tricomi solution of the confluent hypergeometric equation \cite{NIST-book}. In  (\ref{e: left solution}) and (\ref{e: right solution}) the quantities $A$ and $B$ are constants. Using the expression for the Wronskian of the special functions $\chypergf (\alpha ,\gamma;z) $ and $U(\alpha ,\gamma;z)$ ((13.2.34) of Ref.\ \cite{NIST-book}), we obtain that the Wronskian of the solutions $f_a(r,\bar{\omega})$ and $ g_b (r,\bar{\omega})$ is equal to\footnote{As usual we denote the gamma function by $\Gamma(z)$.} 
\begin{align}\label{eq:Wroskiano fg}
W( f_{ a } (r,\bar{ \omega } ), g_{ b } (r,\bar{ \omega } ))= W(\bar{\omega})=-\frac{  AB(2L+1)\Gamma \left( L+ 1 / 2  \right)  }{ 2 \sqrt{ \pi  } {\bar{ \omega }  }^{ 2L+1 }  }.
\end{align}

As we noted previously, for $D$ even the parameter $L$ is an integer, whereas for $D$ odd the parameter $L$ is a half integer. As a consequence the function $g_b (r,\bar{\omega})$ behaves in a different way for $D$ even and $D$ odd. Therefore in what follows we study both cases.

\subsection{D even} \label{ss: D even}

For $D$ even the quantity $L$ is an integer, hence the parameters $\alpha$ and $\gamma$ are integers, thus the Tricomi function that appears in the expression (\ref{e: right solution}) for $g_{ b } (r,\bar{ \omega } )$ takes the form \cite{NIST-book}
\begin{equation}
 U(\alpha ,\gamma; 2 \bar{\omega} r)  = \sum_{s=0}^{L}  \left( \begin{array}{c}
                                                                L \\ s
                                                               \end{array}
 \right) (\alpha)_s \frac{1}{(2 \bar{\omega} r)^{L+1+s}},
\end{equation} 
and the function $ g_{ b } (r,\bar{ \omega } )$ has a branch cut between $\omega=-m$ and $\omega= m$. Therefore the function  $ \tilde{G} (r,r^\prime;\omega)$ has a branch cut in the interval $\omega \in (-m,m)$.

\begin{figure}[ht]
\centering
\includegraphics[scale=1]{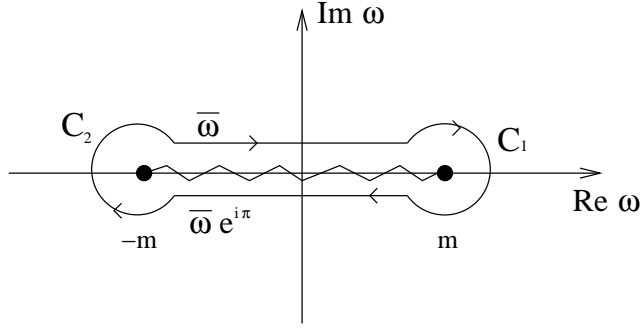}
\caption{Path used to calculate the integral along the cut for $D$ even. We represent the cut by a broken line.}
\label{figure1}
\end{figure}

As usual, to calculate the integral of the formula (\ref{e: Green}) we deform the contour of integration. Thus we close the path of integration with a semicircle of large radius in the lower complex $\omega$ plane. For $t > r + r^\prime$ the integral along this semicircle goes to zero and in our case this condition is satisfied since we are interested in the behavior of the Dirac field as $t \to \infty$.  We notice that in our problem the function $\tilde{G} (r,r^\prime;\omega)$ does not have poles in the complex plane $\omega$. Furthermore it is well known that the late time tail of the field is determined by the branch cut in the integral (\ref{e: Green}). To calculate this integral for $D$ even we use the contour of Figure \ref{figure1} with the branch of $\bar{\omega}$ indicated in this Figure. We obtain that the integral along the cut simplifies to \cite{Hod:1998ra}, \cite{Hod:1997fy}
\begin{align}\label{e: Primera parte integral}
G(r,r';t)  = - \frac{ 1 }{ 2\pi  } & \left[  \int_{-m }^{ m } \right. \frac{ f_a(r',\bar{ \omega } \textrm{e}^{i\pi  }) g_b(r,\bar{ \omega } \textrm{e}^{ i\pi  }) }{ W(\bar{ \omega } \textrm{e}^{ i\pi}) } \textrm{e}^{-i\omega t } d\omega \\
 & -  \int_{ -m }^{ m } \left. \frac{ f_a(r',\bar{ \omega } ) g_b(r,\bar{ \omega } ) }{ W(\bar{ \omega } ) }  \textrm{e}^{ -i\omega t }d\omega    \right] ,  \nonumber
\end{align}
since the integrals along the small circles $C_1$ and $C_2$ of Figure \ref{figure1}  go to zero as their radii go to zero.

For the following computations we need to know the analytical properties of $f_a(r,\bar{ \omega } )$, $g_b(r,\bar{ \omega } )$, and $W(\bar{ \omega } )$ under a rotation $z \to z \textrm{e}^{i \pi}$ of the argument. From (\ref{e: left solution}) we find that the function $f_a(r,\bar{ \omega } )$ fulfills $f_{ a } (r,\bar{\omega} \textrm{e}^{ i\pi  })= f_{ a } (r,\bar{ \omega } )$, that is, this function  is single valued. To evaluate $g_b(r,\bar{ \omega } \textrm{e}^{ i\pi  })$ we use that $U(\alpha ,\gamma; z)$ is given by \cite{Olver-Special}
\begin{align}\label{e: GeneralSen}
U (\alpha ,\gamma ;z) = \frac{ \pi  }{ \sin ( \pi \gamma  )  } \left\{ \frac{ _{ 1 }F_{ 1 }(\alpha ,\gamma ;z) }{ \Gamma (1+\alpha -\gamma )\Gamma (\gamma ) } -\frac{ { z }^{ 1-\gamma  }{_{ 1 }F_{ 1 }}(1+\alpha -\gamma ,2-\gamma ;z) }{ \Gamma (\alpha )\Gamma (2-\gamma ) }  \right\},
\end{align}
and therefore for $D$ even ($L$ integer) we obtain that the function $g_b(r,\bar{ \omega } )$  satisfies 
\begin{align}\label{e: rotation g}
 g_{ b } (r,\bar{ \omega } \textrm{e}^{ i\pi  } )= \frac{ 2B }{ A }  \frac{ (-1)^{ L+1 } \Gamma(L+1) }{ \Gamma(2(L+1))  }  f_{ a } (r,\bar{\omega } ) -  g_{ b } (r,\bar{ \omega } ),
\end{align}
and hence the Wronskian fulfills $W(\bar{ \omega }\textrm{e}^{ i\pi  }) =-W(\bar{ \omega } )$.

From these results  we find that (\ref{e: Primera parte integral}) transforms into 
\begin{align} \label{e: Green second D even}
&G(r,r';t) =-\frac{ 1 }{ 2\pi  } \int_{ -m }^{ m }{ \quad f_a(r',\bar{ \omega } )\left[ \frac{ g_b(r,\bar{ \omega }\textrm{e}^{ i\pi  }) }{ W(\bar{ \omega }\textrm{e}^{ i\pi  }) } -\frac{ g_b(r,\bar{ \omega }) }{ W(\bar{ \omega } ) }  \right] \quad\textrm{e}^{ -i\omega t }d\omega  } \\ 
&=  \frac{ { (-1) }^{ L } }{ A^{ 2 } 2^{2L } \Gamma (L+\frac{ 3 }{ 2 } )\Gamma( L+\frac{ 1 }{ 2 }) (2L+1) }  \int _{ -m }^{ m }{ f_a(r',\bar{\omega } )f_a( r,\bar{ \omega }) \bar{\omega }^{ 2L+1 }\textrm{e}^{ -i\omega t }d\omega  }. \nonumber 
\end{align}
To determine the asymptotic time tail of the Dirac field we must evaluate the previous integral in the limit $\bar{\omega} r \ll 1$ \cite{Hod:1998ra}, \cite{Hod:1997fy}. To make this calculation we use the relationship between the confluent hypergeometric function and the modified Bessel function of first kind $I_\nu (z)$ ((13.6.9) of \cite{NIST-book}) and in contrast to the previous works we consider the first terms of the expansion for $I_\nu (z)$ near $z = 0$. From these facts we obtain that  in the limit $t \gg 1/ m$ the previous integral is equal to
\begin{align} \label{e: integral -m m}
 &\int_{ -m }^{ m } f_a(r',\bar{\omega })f_a(r,\bar{ \omega } ) \bar{\omega }^{ 2L+1 }\textrm{e}^{ -i\omega t }d\omega  = \frac{A^2 \Gamma(L + 3/2)^2 m^{L-1/2} 2^{L+3/2}(r r^\prime)^{L+1}}{t^{L+3/2}}  \\ 
& \times \left[\frac{m}{\Gamma(L + 3/2)} \cos ( \chi_1 ) + \frac{1}{t} \left\{ \frac{4 (L+1)^2 -1}{8 \Gamma(L + 3/2)} + \frac{(r^2+(r^\prime)^2)m^2}{\Gamma(L + 3/2)} \right\} \sin(\chi_1) + \dots  \right], \nonumber  
\end{align}
where
\begin{equation} \label{e: chi angle}
 \chi_1 = m t -\left(\frac{L}{2}+\frac{3}{4} \right) \pi .
\end{equation} 
To calculate the previous integral we use the integral representation of the Bessel function ((9.1.20) of \cite{Abramowitz-book}) and the Hankel asymptotic expansions of the Bessel function ((9.2.5) of \cite{Abramowitz-book}). Thus in the limit $t \gg 1/ m$ the contribution of the cut to the Green function is equal to
\begin{align}
 G(r,r';t) = & \frac{(-1)^L \Gamma(L + 3/2) m^{L-1/2}}{2^{L-3/2} \Gamma(L + 1/2) (2 L +1)}\frac{(r r^\prime)^{L+1}}{t^{L+3/2}}  \left[ \frac{m}{\Gamma(L + 3/2)} \cos ( \chi_1 ) \right. \nonumber   \\
 & \left. + \frac{1}{t} \left\{ \frac{4 (L+1)^2 -1}{8 \Gamma(L + 3/2)} + \frac{(r^2+(r^\prime)^2)m^2}{\Gamma(L + 3/2)} \right\} \sin(\chi_1) + \dots \right].
\end{align}

Notice that we calculate the first two terms of the expansion for $G(r,r';t)$ in inverse powers of $t$. In a similar way we can compute additional terms. Therefore from the previous expression, as $t \to \infty$ the dominant contribution to the function $G(r,r';t)$ is
\begin{align} \label{e: resultado}
 G(r,r';t) & \approx \frac{ \sqrt { 2  } { (-1) }^{ L} { m }^{ L+\frac{ 1 }{ 2 }  }}{{ 2 }^{ L-1 }(2L+1) \Gamma (L+\frac{ 1 }{ 2 } ) }  \frac{{ (r'r) }^{ L+1 } }{{ t }^{ L+ 3 /2   }}  \cos  (\chi_1) .
\end{align}
Thus in this limit the function $G(r,r';t)$  behaves  as
\begin{equation}\label{e: resultado final}
 G(r,r';t) \approx \frac{\cos  (\chi_1)}{t^{L +3/2}} .
\end{equation}

\subsection{D odd} \label{ss: D odd}

We remind that for $D$ odd the parameter $L$ is a half-integer and from the expressions for $\alpha$ and $\gamma$ we obtain that the quantity $\gamma$ is an integer and the parameter  $\alpha$ is a half-integer. Therefore the Tricomi function that appears in the function $ g_{ b } (r,\bar{ \omega })$ is equal to\footnote{As usual we denote the logarithmic derivative of the gamma function by $\psi(z)$.} \cite{NIST-book}
\begin{align}
  U(\alpha ,\gamma; 2 \bar{\omega} r) & = \frac{1}{\Gamma(L+1)} \sum_{k=1}^{2L+1} \frac{(k-1)! (-L+k)_{2 L +1-k} }{(2L+1-k)!} \frac{1}{(2 \bar{\omega}r)^k} \\
  &+ \frac{(-1)^{2L+2}}{(2L+1)! \Gamma(a - 2L-1)} \sum_{k=0}^{\infty}  \frac{(\alpha)_k (2 \bar{\omega} r)^k}{(2L +2)_k k!} \left( \ln (2 \bar{\omega}r) + \psi(\alpha + k)\right. \nonumber \\
  & \left. - \psi(1 + k) - \psi(2 L + 1 + k +1) \right) . \nonumber 
\end{align} 

\begin{figure}[ht]
\centering
\includegraphics[scale=1]{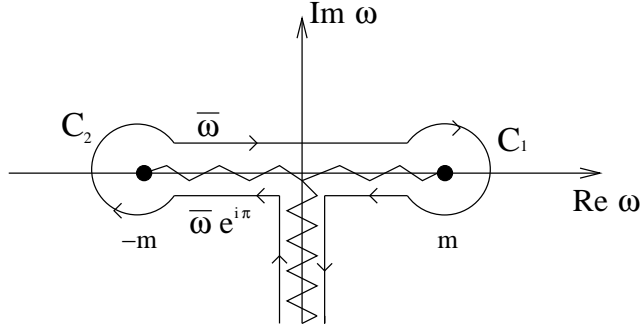}
\caption{Path used to calculate the integral along the cut for $D$ odd. We represent the cut by a broken line.}
\label{figure2}
\end{figure}

As before, for $D$ odd the function $ g_{ b } (r,\bar{\omega})$ has branch points at $\omega = -m$, $\omega=m$. Furthermore the logarithmic term implies that the function $ g_{ b } (r,\bar{\omega})$ has an additional branch point as $\omega \to \infty$. Thus we can choose the branch cut as in Figure \ref{figure2} and using the contour of this Figure we calculate the integral along the branch cut. We notice that the integrals along the two paths parallel to the negative imaginary axis have a sum equal to zero and following a similar method to that for $D$ even we find that for calculating the integral along the cut we must compute an expression similar to that given in (\ref{e: Primera parte integral}). To calculate this integral for $D$ odd we must determine the analytical properties of the functions  $ f_{ a } (r,\bar{\omega})$,  $ g_{ b } (r,\bar{\omega})$, and $W(\bar{\omega})$. For $D$ odd we find that the function $ f_{ a } (r,\bar{\omega})$ is single valued, as for $D$ even, but we find that the function $ g_{ b } (r,\bar{\omega})$ fulfills
\begin{align}\label{e: rotation g D odd}
 g_{ b } (r,\bar{ \omega } \textrm{e}^{ i\pi  } )= \frac{ B }{ A }  \frac{i (-1)^{ L+1/2 } \pi^{1/2}}{ 2^{2 L +1} \Gamma(L+3/2)  }  f_{ a } (r,\bar{\omega } ) + g_{ b } (r,\bar{ \omega } ),
\end{align}
and hence $W(\bar{\omega})$ transforms in the form $W(\bar{ \omega } \textrm{e}^{ i\pi  }) = W(\bar{ \omega } )$.

From these facts  we find that for $D$ odd the analogous of Eq.\ (\ref{e: Green second D even}) is
\begin{align}
G(r,r^\prime;t) =  \frac{ { i (-1) }^{ L+1/2 } 2^{-(2L+1) } }{ A^{ 2 }  \Gamma (L+\frac{ 3 }{ 2 } )\Gamma( L+\frac{ 1 }{ 2 }) (2L+1) }  \int _{ -m }^{ m }{ f_a(r', \bar{ \omega } )f_a(r,\bar{\omega } ) \bar{\omega}^{ 2L+1 }\textrm{e}^{ -i\omega t }d\omega  } .
\end{align}
Using a similar method to that for $D$ even and taking into account the integral (\ref{e: integral -m m}) we obtain that the integral along the cut produces that as $t \to \infty$, the function $ G(r,r';t)$ behaves as
\begin{align}
 G(r,r';t) = & \frac{i (-1)^{L+1/2} \Gamma(L + 3/2) m^{L-1/2}}{2^{L-1/2} \Gamma(L + 1/2) (2 L +1)} \frac{(r r^\prime)^{L+1}}{t^{L+3/2}} \left[ \frac{m}{\Gamma(L + 3/2)} \cos ( \chi_1 ) \right. \nonumber   \\
 & \left. + \frac{1}{t} \left\{ \frac{4 (L+1)^2 -1}{8 \Gamma(L + 3/2)} + \frac{(r^2+(r^\prime)^2)m^2}{\Gamma(L + 3/2)} \right\} \sin(\chi_1) + \dots \right].
\end{align}
Therefore for $D$ odd the dominant term in the late time tail of the Dirac field  is 
\begin{align} \label{e: resultado D odd}
 G(r,r';t) & \approx \frac{i   { (-1) }^{ L+1/2 } { m }^{ L+ 1/2} }{{ 2 }^{ L-1/2 }(2L+1) \Gamma (L+\frac{ 1 }{ 2 } ) }  \frac{{ (r'r) }^{ L+1 } }{{ t }^{ L+ 3/2   }}  \cos  (\chi_1) ,
\end{align}
and hence in the limit $t \gg 1/m$, the function $G(r,r';t)$ behaves as in the formula (\ref{e: resultado final}).

\section{Discussion}
\label{s: Discussion}

As is well known, using $G(r,r^\prime;t)$ we write the solution to Eq.\ (\ref{e: common form F}) in the form \cite{Ching:1994bd}, \cite{Ching:1995tj}
\begin{equation}\label{e: EvolucionCampo}
F(r,t)=\int { dr^\prime G(r,r^\prime;t){ \partial  }_{ t } } F (r^\prime,0) + \int { d r^\prime \partial_{ t }G(r,r^\prime;t) } F(r^\prime,0) ,
\end{equation} 
therefore we find that for $D$ even or $D$ odd and initial data satisfying $F(r^\prime,0)=0$  or $\partial_t F(r^\prime,0)=0$, the Dirac field propagating in the Minkowski spacetime has an oscillatory inverse power law tail of the form
\begin{align} \label{e: first behavior}
F(r,t) \sim \frac{ \cos (\chi_1) }  { { t }^{ L+3 / 2   } } ,
\end{align}
for the first case or
\begin{align} \label{e: second behavior}
F(r,t) \sim \frac{ \sin  (\chi_1)  }  { t^{ L+ 3/ 2 } } ,
\end{align}
for the second case. Thus in the $D$-dimensional Minkowski spacetime as $t \to \infty$ the Dirac field decays with a power $t^{ -(L+ 3/ 2 )}$ and oscillates in the form $\cos (\chi_1) $ or $\sin (\chi_1)$. We point out that these oscillations have a period $2 \pi / m$ and this period is only determined by the value of the mass.

It is convenient to notice that for initial data given by $\partial_t F(r^\prime,0) \neq 0$ and $F(r^\prime,0) \neq 0$, for $t \to \infty$ the behavior  of the Dirac field is
\begin{equation} \label{e: behavior general}
 F(r,t) \sim \frac{1}{{t}^{ L+3 / 2 }} \left[ \tilde{q} \cos (\chi_1)  + m q  \sin (\chi_1)  \right],
\end{equation} 
with
\begin{equation} 
 \tilde{q} = \int dr^\prime (r^\prime)^{L+1} \partial_t F(r^\prime,0), \qquad \qquad q = \int dr^\prime (r^\prime)^{L+1} F (r^\prime,0). 
\end{equation}
This is due to the existence of the factor $\cos (\chi_1)$ in the asymptotic behavior of the Green function  (see the formula (\ref{e: resultado final})). The mentioned factor produces that in the asymptotic limit $G(r,r^\prime;t)$ and $\partial_t G(r,r^\prime;t)$ decay in the form $h(\chi_1)/t^{L+3/2}$ where $h(\chi_1)$ is a sine or cosine function of $\chi_1$.  The behavior (\ref{e: behavior general}) is more complicated than the behavior (\ref{e: first behavior}) or (\ref{e: second behavior}), but usually we consider  the initial conditions $F(r^\prime,0)=0$ or $\partial_t F(r^\prime,0)=0$, that is, the behavior that we find in  (\ref{e: first behavior}) or (\ref{e: second behavior}).

From our results we deduce that  in the $D$-dimensional Minkowski spacetime the Dirac field decays with an inverse power of the time and this power is a half-integer for $D$-even and integral number for $D$ odd. Furthermore we notice that for a  fixed value of $n$, when the dimension of the spacetime increases, the Dirac field decays faster, since the power  $L+3/2$ of the  time increases as $D$ increases (see the expressions (\ref{e: L for F1 F2})). In contrast to massless fields \cite{Cardoso:2003jf}, \cite{Soodak-Tiersten}, for the massive Dirac field our results point out that in Minkowski spacetime its late time behavior is not strongly affected whether $D$ is even or odd, since in both cases we obtain a late time behavior of the form (\ref{e: first behavior}) or (\ref{e: second behavior}).

For the massive Dirac field propagating in four dimensional curved spacetimes, the previous results \cite{Jing:2004xv}--\cite{Moderski:2008nq} show that its asymptotic behavior has two parts (see also \cite{Hod:1998ra}--\cite{Koyama:2001qw}). At the intermediate late time behavior the power law decay rate depends on the angular eigenvalue and the mass of the Dirac field, whereas in the very late time behavior the power law decay rate is of the form $t^{-5/6}$, that is, it is independent of the parameters for the field and the spacetime \cite{Jing:2004xv}--\cite{Moderski:2008nq}. In this work we find that in the $D$-dimensional Minkowski spacetime the asymptotic behavior of the massive Dirac field is different from the two previously described in four dimensional curved spherically symmetric spacetimes, since  as $t \to \infty$ the power law decay rate of the massive Dirac field depends on the dimension of the spacetime, on the mode number $n$ of the angular eigenvalues, and it does not depend on the mass of the Dirac field. Thus in curved spacetimes we can infer that for the massive Dirac field the dependence on the mass of the power law decay rate at intermediate late times is an effect of the curvature of the spacetime.

\section{Acknowledgments}

We thank the support of CONACYT M\'exico, SNI M\'exico, EDI IPN, COFAA IPN, and Research Projects IPN SIP-20150707 and IPN SIP-20151031.

\end{document}